\documentclass[letter,scriptaddress,twocolumn, showkeys]{revtex4}
\usepackage{ amssymb }
	\usepackage{amsmath}
	\usepackage{makeidx}
	\usepackage{amsfonts}
	\usepackage[ansinew]{inputenc}
	\usepackage[usenames,dvipsnames]{pstricks}
	\usepackage{subfigure}
	\usepackage{epsfig}
	\usepackage{pst-grad} 
	\usepackage{pst-plot} 
		\usepackage{amsthm}
\usepackage{ amssymb }
 	\usepackage{makeidx}
	\usepackage{amsfonts}
 	\usepackage{listings}
 	\usepackage{mathtools}
	\usepackage{float}
	
	\usepackage[colorlinks,hyperindex]{hyperref}
	\hypersetup
	{
		colorlinks,%
		citecolor=blue,%
		linkcolor=blue,%
		urlcolor=blue,%
	}
	\usepackage{tikz}

\usepackage{amsmath}
\usepackage{amssymb}
\usepackage{graphicx}
\usepackage{xcolor}
\usepackage{float,graphicx}
\usepackage{placeins}
\usepackage{color, colortbl}
\usepackage[colorinlistoftodos]{todonotes}
\usepackage{array}

\theoremstyle{definition}

	\theoremstyle{plain}

\makeindex

\begin{document}

\title{Housing Forecasts via Stock Market Indicators}

\author{Varun Mittal$^{a}$ and  Laura P. Schaposnik$^{\star, b}$, 
}
  \affiliation{($\star$) Corresponding author: schapos@uic.edu}

\begin{abstract}
Through the reinterpretation of housing data as candlesticks, we extend {\it Nature Scientific Reports} article by  Liang and Unwin [LU22] on stock market indicators for COVID-19 data, and utilize some of the most prominent technical indicators from the stock market to estimate future changes in the housing market,  comparing the findings to those one would obtain from studying real estate ETF's. By providing an analysis of MACD, RSI, and Candlestick indicators (Bullish Engulfing, Bearish Engulfing, Hanging Man, and Hammer), we  exhibit their statistical significance in making predictions for USA data sets  (using Zillow Housing data) and also consider their applications within three different scenarios: 
a stable housing market, a volatile housing market, and a   saturated market. In particular, we show  that bearish indicators  have a much higher statistical significance then bullish indicators, and we further illustrate how   in less stable or more populated countries, bearish trends are only slightly more statistically present compared to bullish trends.  
 \end{abstract}

\maketitle
 
\section{Introduction}
The mathematical study of housing markets dates back to the 1970s \cite{10.2307/621308}, and its   prominence has seen a significant increase in the last decades, as consumers are increasingly looking at the right time to make a significant investment \cite{GONG201951}. Whilst classical studies have included complex models and economic examinations to varying levels of success \cite{doi:10.1080/00420988920080181}, after the housing market crashed in 2008 it  became clear that new models are necessary to understand both the local and global behaviour of housing markets. Since then, more novel approaches begun to be implemented \cite{6073000}, but these are yet to become as robust or successful as they could potentially be \cite{doi:10.1068/a030243}. 
Purchasing a house is never easy, and buying at the optimal time is a significant decision. As a result of COVID, the housing market has become more volatile and has fluctuated significantly \cite{LIU2021110010}. Therefore, we propose the utilization of technical indicators that are commonly applied in the stock market to the housing markets to identify both bearish and bullish trends in the housing market, where

\begin{itemize}
    \item[(a)] 
    a trend is considered \textbf{bullish} if it is corresponds to an increase in value for a time series. 
    \item[(b)]  
    a trend is considered \textbf{bearish} if it corresponds to a decrease in value for a time series.
    \end{itemize}


In a recent {\it Nature Scientific Reports} article,   Liang and Unwin proposed to use stock market indicators to understand trends in COVID-19 cases in the US  \cite{liang_unwin_2021}. Inspired by this work,  we shall consider Zillow's data set of the USA's housing market and study it through examining the usability of statistical indicators usually seen in the stock market such as  MACD Bearish, MACD Bullish, RSI Bearish, RSI Bullish, Hanging Man, and Hammer. These indicators can be used to identify trends (both bearish and bullish) that can be vital when it comes to house purchasing decisions. In what follows we shall show via these indicators, trends (both bearish and bullish) can be detected and used to understand housing prices. 

For each indicator  we calculate the statistical significance using the Wilcoxon test, and show in particular  that while both RSI indicators were statistically significant, only the bearish indicator is significant for MACD. Finally, we extend our analysis to other countries, with the intent of being able to make insights in various types of markets. To be encompassing and robust, we considered   different types of markets (Saturated, Volatile and Stable) and studied the relevance of the indicators within each setting.  In particular, we show that
\begin{itemize}
\item   across all countries, bearish indicators were much more statistically significant compared to bullish indicators, for the vast majority of indicators..
\item  the difference in the statistical significance of bullish and bearish trends isn't nearly as large in saturated and volatile housing markets. 
\item when using sock market indicators directly on housing prices, the trends of the housing price can be predicted more accuratedly than when considering the indicators on housing ETFs. 
\end{itemize}



To carry out our  study, we considered RSI, MACD, and Candlestick Analysis conducted through R. After grouping  grouping Zillow's data \cite{zillow} into  Heikin Ashi candlesticks, we scanned the data for the four candlestick indicators mentioned above and  used the Wilcox, and later performed a similar procedure for the other indicators.     
  After introducing some background in Section \ref{machine}, we dedicate Section \ref{results} and Section \ref{conti} to the main findings of our work, which  can be seen in two different directions in terms of the indicators themselves and their statistical significance in various markets. To highlight the utility of our approach vs any standard stock market study of Housing ETFs, we compared our findings to what one can deduce through the same indicators on the most used Housing ETFs, namely  VNQ	(Vanguard Real Estate ETF), SCHH	(Schwab US REIT ETF), and 	 XLRE	(Real Estate Select Sector SPDR Fund) in Section \ref{etf}.  Finally, we expand on the analysis and applications of the above findings in Section \ref{last}. Finally, we also attached our code for the project on Github \cite{varun121322_2022}.

\section{Background: Stock market indicators}\label{machine}

Researchers have considered a large number of ways that stock market prices can be forecasted, for example by speculating a stock's intrinsic value, employing complex algorithms, and speculating human behavior \cite{CHAUVET200087}, or by using some  complex algorithms such as rolling window analysis and deep learning \cite{https://doi.org/10.48550/arxiv.1909.10660}. Moreover, recently there has been some success around more simple but nuanced indicators, popularly known as candlestick analysis, MACD, and RSI \cite{jrfm7010001}. Stock indicators are being used increasingly to identify trends in the stock market, as traders are looking to maximize their profits by trying to gain insight on the future of the market.

The current research is inspired by the  work in \cite{liang_unwin_2021} where stock market indicators were used to forecast COVID-19 cases, and thus we shall dedicate this section to review some of the main results which shall prove useful for our research and for comparison with our own findings.    The indicators considered in  \cite{liang_unwin_2021} were candlestick analysis, RSI and MACD indicators. After giving a mathematical definition of  those indicators for a time series of datapoints, the authors measured their statistical significance   by first considering the null hypothesis, the probability that an external circumstance was the cause of a bearish or bullish trend, not an indicator (recall that an indicator is described as \textbf{bullish} if that indicator predicts a future decrease while an indicator is described as \textbf{bearish} if it describes a future increase in prices). This probability was denoted as a $p$-value, and it was then ascertained that lower $p$-values  translated to higher statistical significances as that meant the probability of a trend being forecasted by an indicator was higher in comparison to a random external event. After considering this hypothesis, they used the Wilcoxon test for R to determine these $p$-values  across all of the indicators. This allowed them to make conclusions about the applicability of indicators. 

In what follows we will give a brief overview of the work in \cite{liang_unwin_2021} on candlestick analysis, RSI and MACD indicators introduced for the timed datasets of COVID-19 cases, and  we will then extend to the timed housing data of \cite{zillow}. 

\subsection{Candlestick Analysis}
The first type of indicators considered were candlestick indicators, which included Hammer, Bullish Engulfing, Dark Cloud Over, Bearish Engulfing, and Hanging Man.

\begin{figure}[H]
  \centering \includegraphics[scale= 0.12]{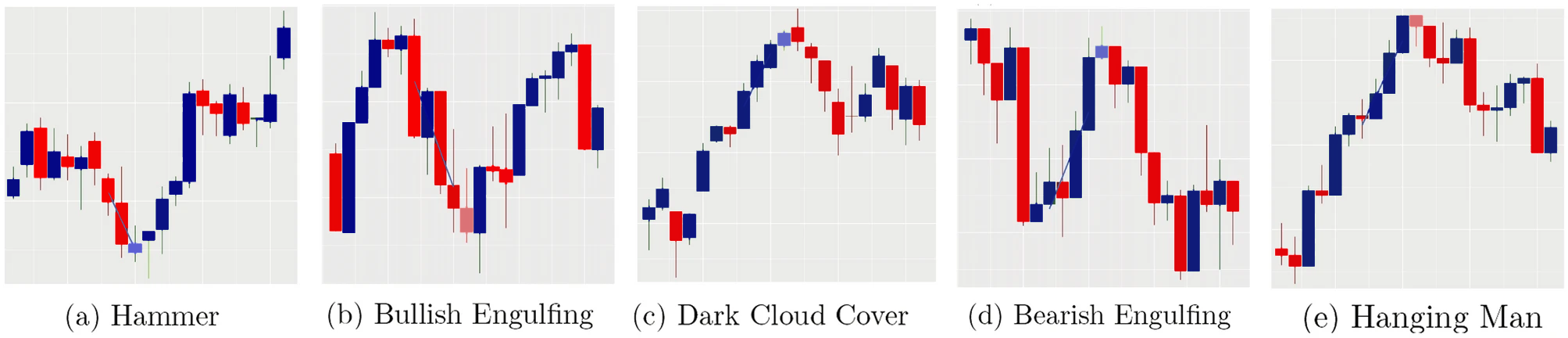}
  \caption{Candlestick patterns studied in \cite{liang_unwin_2021}.}
      \label{James1}
\end{figure}

Weekly candlesticks were constructed with the real body constituting the region marked by the endpoints of the opening and closing values (covid cases), and the upper and lower limits being the maximum and minimum cases over the time interval respectively. Then, the authors developed R code was developed to identify the candlestick patterns shown in \autoref{James1}  within the dataset to identify the frequencies and predictive capabilities of each of these patterns occurred.
\subsection{RSI}
The RSI (Relative Strength Index), is a popular indicator that is similar to considering the recent ``momentum" of a time series through average rate of increases and decreases in a market, in this case the market in consideration (housing, COVID, or stock). The indicator is constructed by dividing the closing values ${C_{t}}$
over some period into two sets:
\begin{itemize}
    
    \item
    The set ${G_{t}}$ in which the series increased:
    $${G_{t}} = \frac{C_{t} - C_{t - 1}}{C_{t}}, ~{C_{t}}>{C_{t - 1}}.$$

    \item  
    The set ${D_{t}}$ in which the series decreased:
    $${D_{t}} = \frac{C_{t - 1} - C_{t}}{C_{t}}      ~{C_{t}}<{C_{t - 1}}.$$
    \end{itemize}
From the above sets one can compute the averages $A_{t}$ and $B_{t}$ using the EMA $V_{n}$ over $n$ periods with a smoothing factor of $\frac{1}{n}$ leading to
\[A_{t} = V_{n}[G_{t}]~ {\rm and} ~{D_{t}} = V_{n}[D_{t}].\]
Then, the RSI$_{t}$ indicator at time $t$ is defined as follows
\[{\rm RSI}_{t} = 100 - \frac{100}{1 + \frac{G'_{t}}{D_{t}}}\]

Within the stock market, the indicator becomes useful when the RSI value exceeds a lower and upper threshold, as that is when the stock in consideration becomes overbought or oversold, as seen in \autoref{rsiStock}. If the RSI value falls below the lower threshold, that signifies a stock being oversold, which signals a bullish trend as stock prices will soon increase in price. Conversely, if the RSI value exceeds an upper threshold, that signifies a stock being overbought, signaling a bearish trend as stock prices will fall. \begin{figure}[H]
  \centering \includegraphics[scale= 0.6]{image1.png}
  \caption{RSI indicator on Zillow Stock.}
      \label{rsiStock}
\end{figure}

\subsection{MACD}
The MACD indicator (Moving Average Convergence Divergence) is based on a moving average and a signal line. In  MACD and signal line values for the weekly candlesticks. By considering weekly COVID-19 cases, \cite{liang_unwin_2021} identified when the MACD values crossed the signal values from above (bearish trend) and when the MACD values crossed the signal lines from below (bullish trend), and calculated the statistical significance of both of these indicators. The MACD is obtained using two exponential moving averages (EMAs), calculated over two periods of differing length $n$. Specifically, for a given dataset of length $n$, labelling  the closing values  by $C_{i}$, the EMA $V_{n}$ is calculated recursively as
\begin{eqnarray}
V_{i}[C_{i}] =&           C_{1} \quad ~&\text{if} ~ i = 1; \nonumber\\
      =&    sC_{i} + (1 - s)V_{i - 1} \quad ~&\text{if} ~ i > 1, \nonumber
\end{eqnarray}
where S = $\frac{2}{n+1}$ is a smoothing factor. Thus, $V_{n}$ can be seen as the exponential average over $n$ intervals, which by substitutions can be expressed 
\[V_n = R[C_{n} + (1 - R)C_{n-1} \ldots (1 - R)^{n-1}C_{1}]\]
Note that the coefficient of each term decreases exponentially for earlier values in the time series, thus giving
greater weighting to more recent data, hence the name.
Given the EMA, the MACD is defined as the difference
between a longer period average $n_{2}$ and a shorter period average $n_{1}$ (where by convention $n_{1} < n_{2}$) as follows
\[{\rm 
MACD}_{(n_{1}, n_{2})} = V_{n_{1}} - V_{n_{2}}\]

\autoref{macdStock} shows the MACD indicators in action. As depicted in the graphs a bullish trend with MACD occurs when the MACD line crosses the signal line from below, and a bearish trend within MACD occurs when the MACD line crosses the signal line from above. 
\subsection{Statistical Significance}
As mentioned above statistical significance values were calculated in  \cite{liang_unwin_2021} using the Wilcoxon test for R, after consideration of the null hypothesis. Through this robust test to measure the significance (due its non-reliability on normalization and it is a non-parametric method) the authors showed that   both bearish and bullish indicators were significant in consideration of COVID cases. Moreover, whilst their study highlighted that the  3 bullish trends given by Hammer, Bullish Engulfing, and Bullish MACD, and  bearish trend given by Bearish MACD were statistically significant, they shoed that  with high $p$-values, were determined to be statistically insignificant in forecasting COVID cases.
 
 Along the same lines of research in what follows we shall consider  housing market trends through the above indicators, as well as through  the Heikin Ashi Candlestick indicator, which we shall introduce and do a statistical study of in the upcoming sections.  



%

 \section{Heikin Ashi candlesticks}\label{hCandles}
In order to take into account individual data points that alter the statistical significance levels of an indicator, and hence to be able to notice  trends more clearly, it is sometimes useful to consider {\bf Heikin Ashi candlesticks}. These candlesticks are similar to the normal candlestick described before, except  that the individual bars are calculated differently by defining the following:   \begin{eqnarray}
{\rm Close}_n &:=& 1/4[{\rm Open}_n + {\rm High}_n + {\rm Low}_n + {\rm Close}_n]; \nonumber \\
  {\rm Open}_n &:=& 1/2[{\rm Close}_{n - 1} + {\rm Open}_{n - 1}]; \nonumber\\
{\rm High}_n &:=& Max[{\rm High}_n, {\rm Open}_n, {\rm Close}_n];\nonumber\\
{\rm Close}_n &:=& Min[{\rm High}_n, {\rm Low}_n, {\rm Close}_n].\nonumber
\end{eqnarray}

\bigskip
It is important to notice how the terms for   calculating the open values are recursively taking the mean
of the previous open and close values. This lends itself to better forecasting because it signifies a continuity within a trend. For example, with a normal candlestick analysis, there can be certain days that contradict the trend, which makes analyzing and identification of trends difficult (e.g. see third blue candle surrounded by red candles in \autoref{rsiStock}).  When using Heikin Ashi candlesticks, these values are being recursively calculated, and thus it has the effect of smoothing values over time, such that trends will persist and be more apparent for analysis. This solved our problem of the contradictory points, and can also be compared to the smoothing factor in   \cite{liang_unwin_2021} in the mathematical expression of the RSI indicator. 
 
\begin{figure}[H]
  \centering \includegraphics[scale= 0.33]{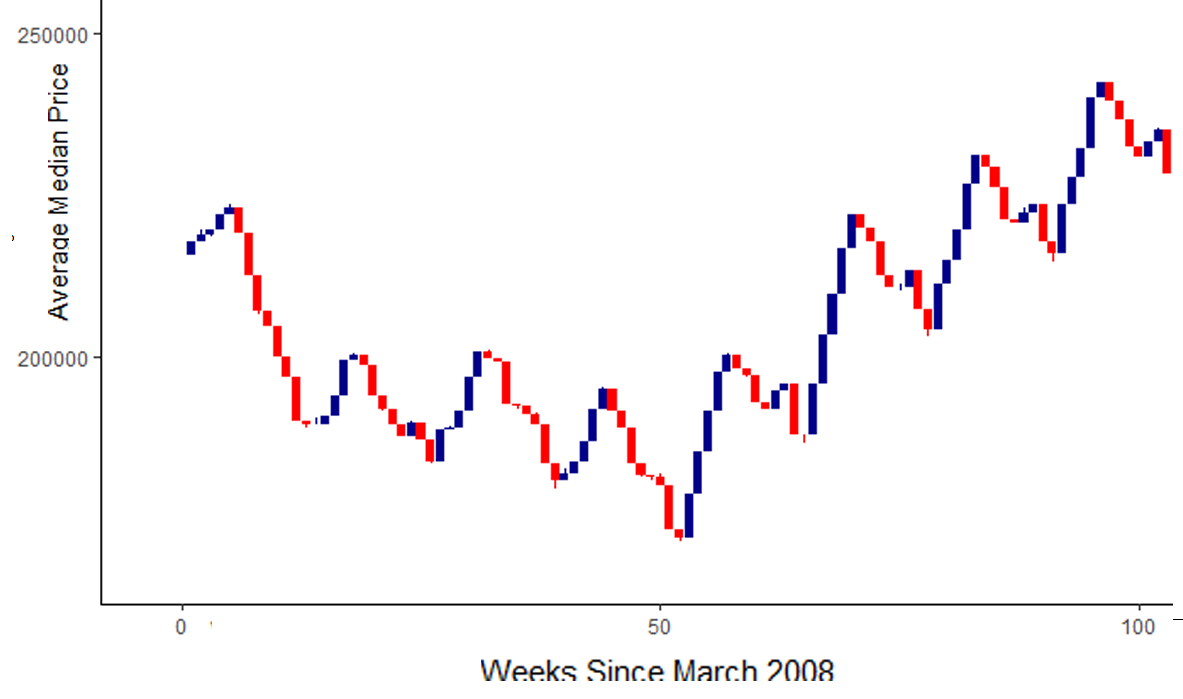}
  \caption{Heikin Ashi candlesticks for Housing Prices}
      \label{housingPricesNormCandles}
\end{figure}

In summary, normal candlesticks can fluctuate more within a certain trend, however, since Heiken Ashi relies on averages, trends are more continuous and observable for analysis as shown in \autoref{housingPricesNormCandles}. This lends itself better to analysis and heightens the statistically significance of indicators, since it makes it more clear that indicators were more likely the sole thing connected to a change, rather then a random external factor.    

The statisical significance is assisted because it centralizes around
 a $p$-value, or the probability that a random external event is the cause of a trend occurring. Therefore, by eliminating small contradictory points, the potential of the $p$-value being large is made smaller as it isolates the trend as causing the change. 


 \section{Housing Data vs ETF Data}
 
 In order to demonstrate the utility of stock market indicators for analizing housing data from \cite{zillow}, we shall consider here another type of data driven from housing prices which one could use to forecast housing market trends, which is given by housing  Exchange-Trade Funds (ETF's).  These are a popular type of pooled investment security that can be purchased or sold on a stock exchange similar to a normal stock, but which is formed from a weighted set of many stocks. To compare the analysis of house prices from \cite{zillow} and ETF's, we shall consider here {\it Vanguard Real Estate Index Fund ETF} (VNQ) because of its high AUM (Assests Under Management) and relative prominence in the market, for which a candlestick chart appears in   \autoref{etfCandles} below.  
\begin{figure}[H]
\hspace{-0.2 in}
  \centering \includegraphics[scale= 0.47]{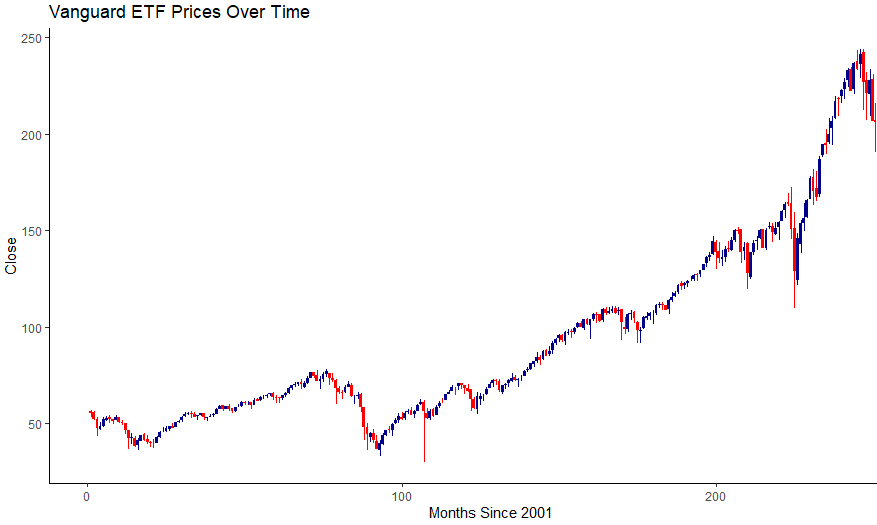}
  \caption{A visualization showing a candelstick chart of Vanguard Real Estate Index Fund ETF
 since 2001. Similar to housing prices, this real estate ETF has increased over time. }
      \label{etfCandles}
\end{figure}

Since ETFs are easier to analyze and study than larger datasets, it is natural to wonder whether studying ETFs alone (without considering housing prices as our study did) would be sufficient for accurate trend prediction.   To study this, we have considered the stock market indicators described in Section \ref{machine} to VNQ's candlestick chart. 
 \begin{figure}[H]
  \centering \includegraphics[scale= 0.25]{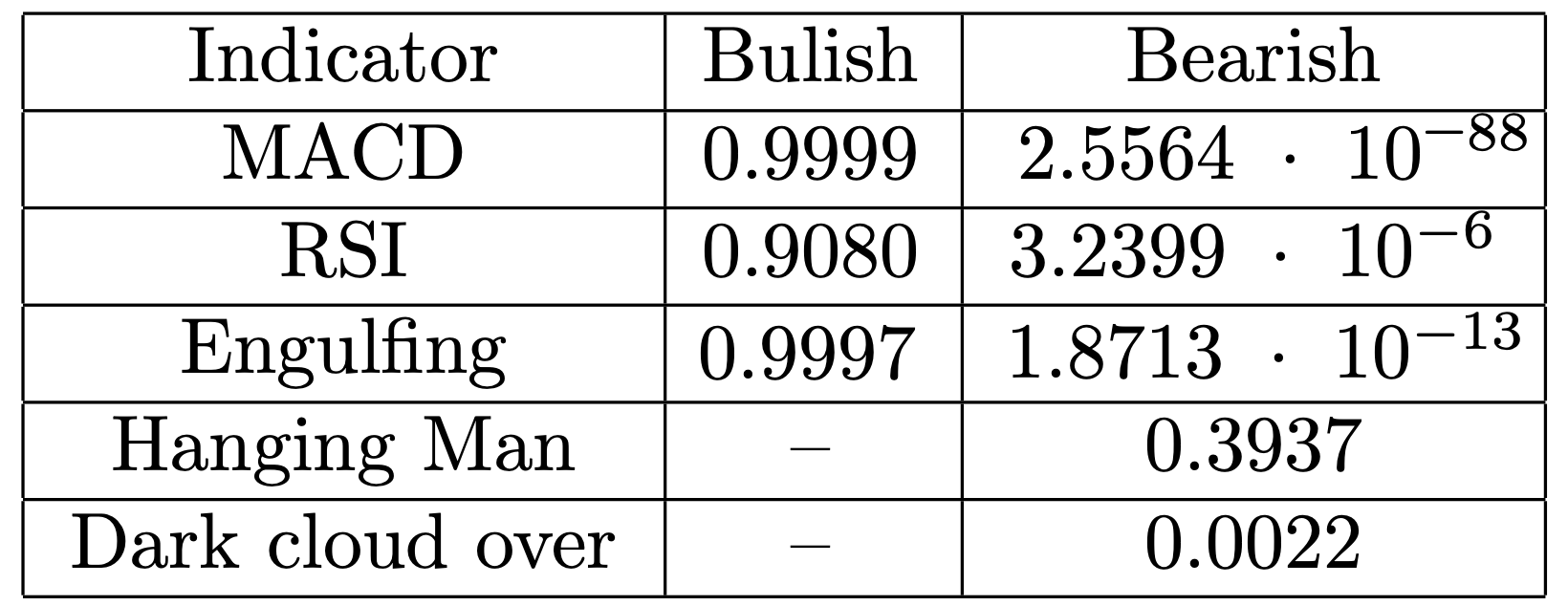}
  \caption{$p$-values  for the different indicators when applied to VNQ's candlestick chart.}
      \label{etfTable}
\end{figure}

To understand the relevance of Zillow's dataset \cite{zillow} we studied the  the statistical significance of the different stock market indicators considered in our paper when applied to VNQ's candlestick chart, for which a table of $p$-values  appears in  \autoref{etfTable} (the probability that a external factors caused the reversal being studied, and not the indicator -- more on this  in Section \ref{results}). 

These probabilities were calculated with the Wilcoxon test in R, after   scanning the code for each of these indicators (MACD, RSI, and Candlestick Indicators which included Bearish and Bullish Engulfing, Hanging Man, and Dark Cloud Over) -- the definition of these indicators is provided in Section \ref{machine}.    
In particular, recall that  since we want to observe the efficacy of indicators,  high statistical significance is associated with a low $p$-value for any indicator. As shown in   \autoref{etfTable},  all of the bearish indicators have low $p$-values  (signifying high statistical significance) while all of the bullish indicators have high $p$-values  (signifying low statistical significance). As a result, one can see that studying ETFs alone would not lead to enough insight into predicting housing market trends since bullish trends could not be foreseen with the usage of technical indicators.
 
From the above analysis, one can see that it becomes of much interest to perform the study of stock market indicators directly on housing data, and thus we shall dedicate the following sections to do so by considering actual housing market data for countries across the globe within varying housing markets, including saturated, volatile, and stable housing markets. Our full results, summarized in Section \ref{results}, can be seen with   \autoref{summaryTable1}. In particular, we shall demonstrate how some bullish indicators have low $p$-values  (correlating to a high statistical significance), for example MACD Bullish in Canada has a $p$-value of 0.11, implying that more insights could be made with bullish indicators through the consideration of housing data rather then ETF prices.

    \section{Housing market trends}\label{results}
\begin{figure}[H]
  \centering \includegraphics[scale= 0.34]{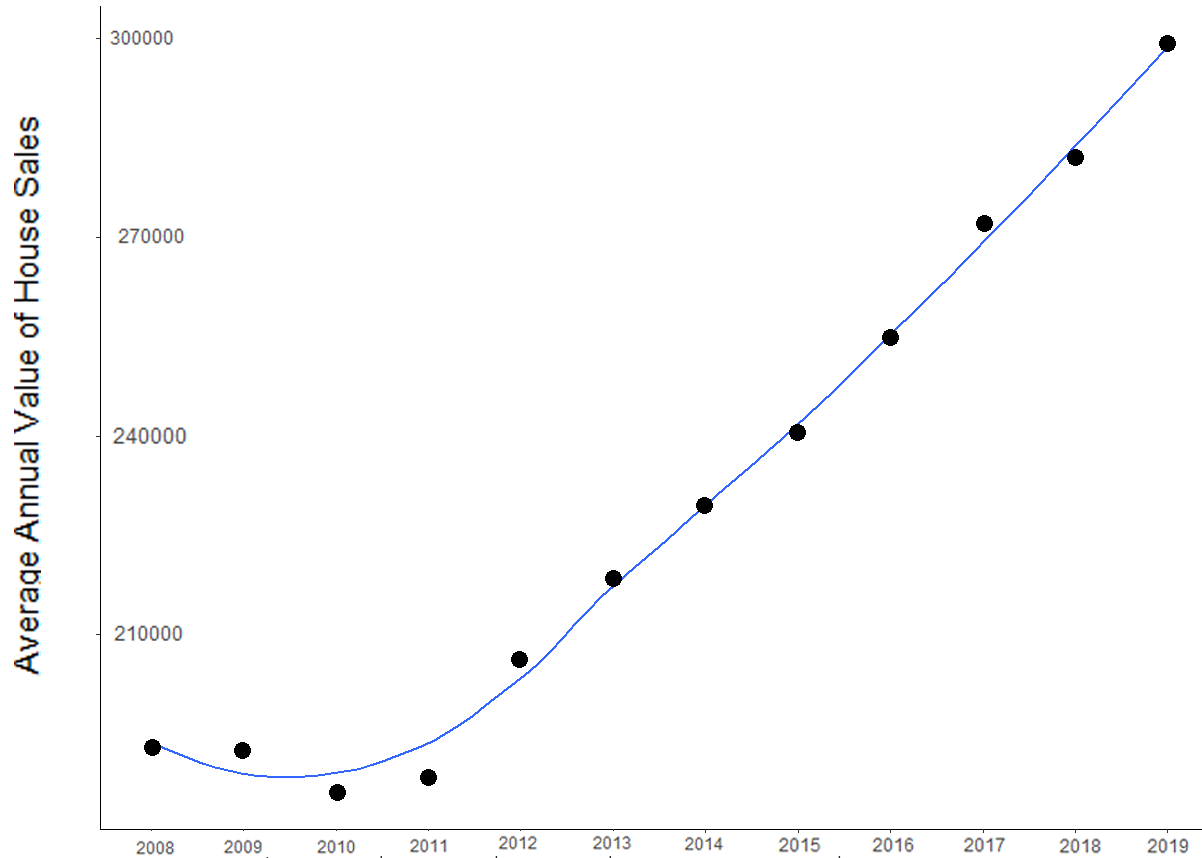}
  \caption{Housing Prices Over Time (original data graph) plotted through Zillow's dataset \cite{zillow}.}
      \label{originalHousingPrices}
\end{figure}

    In order to understand housing market trends via stock market indicators, we shall consider Zillow Data set \cite{zillow} and the information presented there on  average median housing sales per week in a region. The data used together with an approximation through a lienal regression are shown in \autoref{originalHousingPrices}.

In order to compare our study of the US housing market with global markets, we shall consider in this paper the following three different types of housing markets: 

\begin{itemize}
    \item[(a)] 
    A \textbf{saturated} housing market has a lot of activity and is usually correlational with a large population. These markets, with their large-scale activity, have high volume, and generate large numbers of transactions from both purchasers and sellers.
    \item[(b)]  
    A \textbf{volatile} housing market is one that changes quite frequently. For one period of time, it may be undergoing a bearish trend, but then it undergoes a sudden reversal in a short span of time and it is now growing in value. One purpose of our study is being able to forecast these reversals with technical indicators typically found in the stock market. 
    \item[(c)] 
    A \textbf{stable} housing market is one that has sustained trends and doesn't plunge or proliferate suddenly in value across a time period.
    \end{itemize}

We shall study the above markets by considering representatives from each set, and look at the housing data from  Germany (a stable housing market), Canada (a volatile housing market), and Japan (a saturated housing market), thus allowing for our analysis to  yield powerful insights into not one but multiple different markets.
     In order to understand which indicators can be used within each setting, we considered the significance of all of our indicatorsby using  the Wilcoxon test to measure the statistical significance of our indicators.
     
     Following standard conventions, we  considered the null hypothesis being that a trend occurred randomly rather than an indicator inducing it. We called this probability the $p$-value, so the larger the $p$-value, the less statistically significant our indicator is -- and  the closer the $p$-value was to 0, the more statistically significant our indicator is, since there is less of an inherent random event changing the housing prices compared to the indicator taking effect. 
     
\begin{figure}[H]  \centering \includegraphics[scale= 0.37]{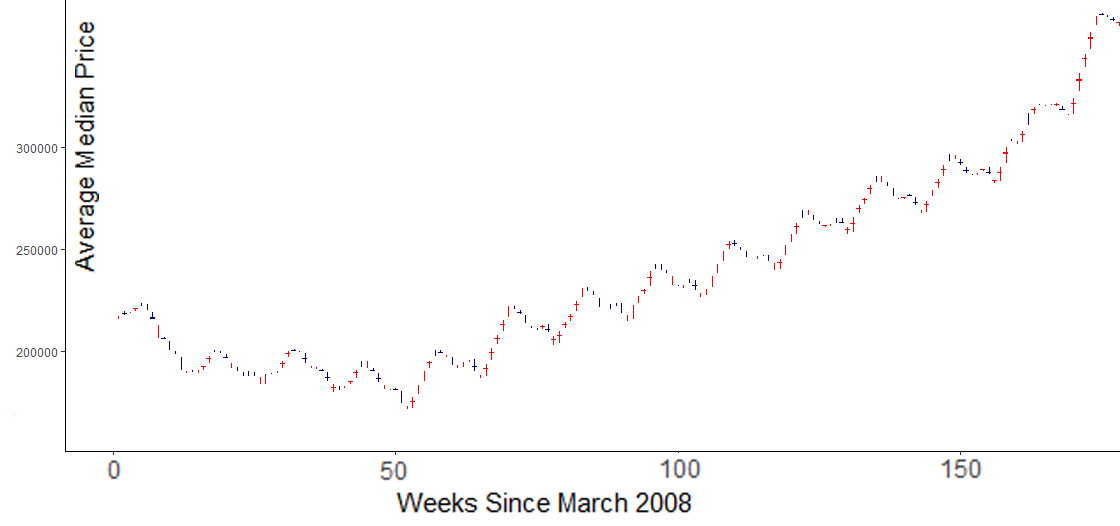}
  \caption{Housing Prices Over Time Heikin Ashi Candlesticks, obtained via geom$\_$candlestick() function in R with appropriate candlestick data. }
      \label{housingPricesHeikinAshi}
\end{figure}

 We also utilized historical housing data from Germany, Canada, and Japan \cite{fred}.  We grouped our data in the form of monthly Heikin Ashi candlesticks for the Zillow (USA) data as shown in \autoref{housingPricesHeikinAshi}, and then yearly candlesticks for the various countries and utilized these  datasets to substantiate and support our claims. 
       
\subsection{Predictions via RSI}
\begin{figure}[H]
  \centering \includegraphics[scale = 0.45]{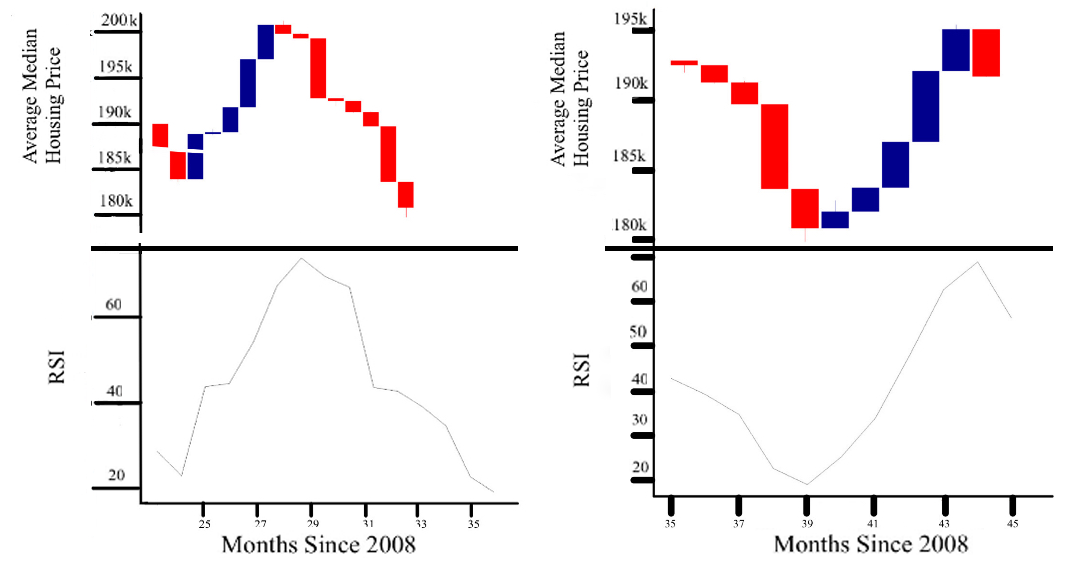}
  \caption{On the left,  RSI value crosses an upper threshold, corresponding to a bearish trend, or a decrease in housing prices. Conversely,  on the right the RSI value crosses a lower threshold, corresponding to a bullish trend, or an increase in housing prices. }
      \label{nonConclusionRSI}
\end{figure}

In what follows we shall consider the RSI indicator on the Zillow Data Set \cite{zillow} as a way to forecast trends, both bearish and bullish. We developed R code to produce RSI values of monthly candlesticks, and observed viable lower and upper thresholds respectively. We then produced a model that would identify the frequency of both RSI Bearish and Bearish indicators, noticing two interesting trends:
    \begin{itemize}
    \item 
   Except for the US housing market with Zillow Data, all RSI Bearish indicators were not statistically insignificant, with $p$-values  larger then 0.9.
    \item 
    When considering the indicator on all types of markets, RSI is the only non-candlestick indicator across all 4 countries in examination that had a statistically significant bearish indicator. Moreover, bullish indicators were much more statistically significant than bearish trends.
   \end{itemize} 
   The RSI indicator  for the Zillow dataset of \cite{zillow} is shown in  \autoref{nonConclusionRSI}:  an example of a bearish trend is indicated by the RSI value crossing an upper threshold (as in the left visualization), and a bullish trend is indicated by the RSI value crossing a lower threshold (as in the right image).

   \subsection{Predictions via MACD}\label{macdPredict}
 
\begin{figure}[H]
  \centering \includegraphics[scale = 0.35]{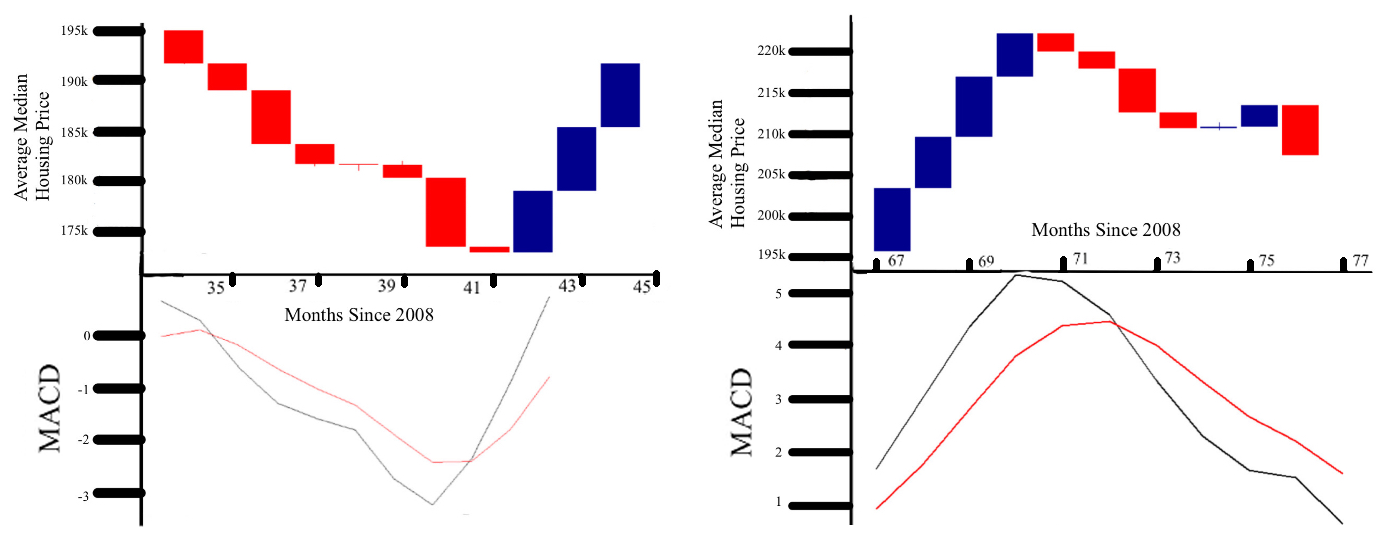}
  \caption{ MACD indicator on Zillow (USA) housing  data.
 }
      \label{nonConclusionMACD}
\end{figure}

The MACD indicator is obtained via a correlation of the moving average and a signal line obtained through the timed data series. By developing R code that produces MACD and signal lines for Zillow's dataset \cite{zillow}, we could combine it with the visualization via monthly candlesticks. When the MACD crossed the signal line from above it signified a bearish reversal,  and when it crossed  from below it signified a bullish reversal, and this is  shown in \autoref{nonConclusionMACD}, where a bearish trend can be seen in the left visualization, and a bullish trend   in the right visualization. 

As before, $p$-values  were calculated for each indicator using the Wilcoxon test over monthly  intervals as described above.
From our study we see different levels of accuracy obtained in our predictions:  
    \begin{itemize}
    \item 
    MACD signals were the most frequently occurring signals that our R code detected. This is salient since it might be the most ubiquitous for consumers to utilize for their benefit.
    \item 
   The  statistical significance of the MACD Bearish and MACD Bullish indicators is very large: whilst MACD Bullish has a low $p$-value with high statistical significance ($7.384 * 10^{-15}$), MACD Bearish has a high $p$-value with low statistical significance (0.99). This was also observed for the  RSI values, with the exception being the US housing market.
   \end{itemize} 
             
    \subsection{Predictions via Heikin Ashi}\label{hCandles}
   We examined the dataset with respect to 4 different candlestick patterns via Heikin Ashi candlesticks, Hammer, Hanging Man, Bearish Engulfing, and Bullish Engulfing. We then scanned the data with R code for these indicators, and then used the Wilcoxon test reiterated above to determine the statistical significance.
    From our study we saw different levels of accuracy obtained in our predictions, allowing us to infer certain candlestick pattern indicators were more viable with respect to others:
    
    \begin{itemize}
    \item
Both within the US and as well as for the other housing markets,  bearish trends were less statistically significant compared to bullish trends.
    \item 
    Within the US, as with the previous indicators,  the Bearish signal (Hanging Man) is determined to be statistically insignificant with a $p$-value of 0.97, but the Bullish signal is determined to be a viable indicator and statistically significant indicator with a $p$-value of 0.05. 
    \item Across the other markets studied,  the converse was observed: $p$-values  for bearish indicators were lower (high statistical significance) and the $p$-values  for bullish indicators were higher (low statistical significance). A table of statistical significance for all of the indicators is shown in \autoref{summaryTable1} below, and will be the focus of our analysis in later sections.
   \end{itemize} 
   
    \begin{figure}[H]
  \centering \includegraphics[scale = 0.65]{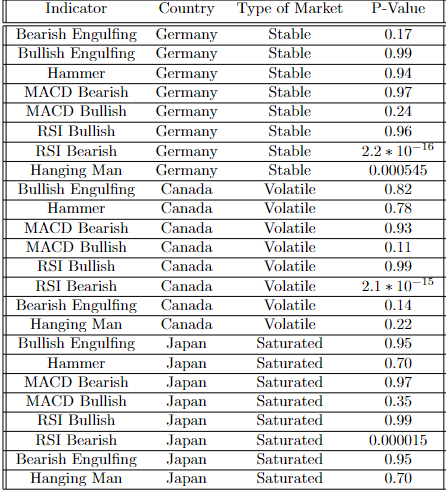}
  \caption{A table of all of the determined $p$-values, broken down by country and market that underwent analysis. }
      \label{summaryTable1}
\end{figure}

    \section{Discussion on Housing market via Stock market indicators}\label{conti}

Having determined all of the $p$-values  and the statistical significance values for the indicators considered in the present paper (see \autoref{summaryTable1}), we shall describe here several interesting trends that provide  insight into the behaviour of the US hosing market as well as the three different global housing markets we considered. For context, recall that we selected and analyzed the US housing market, the German housing market (representative of a stable economy), the Canadian housing market (representative of a volatile economy), and the Japanese housing market (representative of a saturated market).

Through our analysis of the MACD indicator on housing data, and in accordance with \autoref{summaryTable1}, one could observe the following:\begin{itemize}
\item   
This indicator followed the pattern that is emerging within our analysis that exhibited high statistical significance for bullish trends and low statistical significance for bearish trend, but only for the United States.
\item
For other countries, the above scenario was reversed:  bearish MACD indicators were more statistically significant compared to bullish trends.

\end{itemize}
From the above, one can see that in particular while MACD Bullish can be used by both consumers and sellers in the housing market, people who are looking to forecast bearish reversals in the housing market should not look to utilize this indicator.
But the most important thing to realize is the difference in the statistical significance between the MACD Bearish and Bullish.

 In more populated, saturated, and volatile markets, the difference in statistical significance of the MACD Bearish and Bullish is smaller, as exemplified by Japan. In Japan, the difference between MACD Bearish and Bullish is 0.62, but in Germany is significantly larger, hovering around 0.73. This variance in difference is a salient insight because it exhibits that indicators, both statistically significant and insignificant, can be evaluated differently. In a market with a large difference, the failures and effectiveness of indicators can be more established (due to the higher polarization of $p$-values), but this is not  the case with more inherent randomness with a smaller difference and decreased polarization. Moreover,  this trend is persisting when the significance of the MACD is at a constantly low value ($p$-value $>$ 0.9). This signifies that MACD Bearish trends are less statistically significant in more crowded and volatile housing markets, something that could provide   important information for consumers in that region looking to save money. 
         
 From a different perspective, the RSI indicator  proved to uphold the trends we observed throughout our examination. This trend, as reiterated above, is that bullish indicators were less statistically significant in comparison to bearish indicators, across all three housing markets (studied through their representative countries),  except for the United States. 
 
 In the United States, as exhibited by the table in \autoref{summaryTable1}, RSI is very significant statistically in terms of both bullish and bearish. The United States case is interesting because the US can be justified both as a populated yet stable market, so further research into the US housing market can lend itself to even more intriguing analysis. From our above analysis, the following main takeaways are important:
 
\begin{itemize}
\item   
For the United States, RSI is the only statistically significant and viable indicator for bearish trends. This is important for buyers who are looking to purchase at the optimal price.
\item
The $p$-values  for RSI were mostly similar across the three markets under consideration, which signifies that the circumstances of a market may not dictate RSI's efficacy. This makes sense considering RSI's emphasis on recency over long term processes.
\end{itemize}

Given the above analysis, it seems almost contradictory that the RSI indicators can have a low statistical significance. After all, RSI's focus on recency and its practicality when identifying trends should lend itself to higher statistical significance. However, this is where one needs to remember that external circumstances still must be considered, as not all markets are conducive to high statistical significance (low $p$-values).

 Finally, the most appealing analysis is the one that can be done Heikin Ashi candlesticks:
 \begin{itemize}
 \item  Similar to RSI and MACD and as exemplified in the table of  \autoref{summaryTable1}, the bearish indicators (Hanging Man, Bearish Engulfing) were more statistically significant then the bullish indicators (Hammer, Bullish Engulfing). 
 \item The above   trend diddid not continue in the housing markets for Japan and the US, which constituted our saturated/popular housing markets in our study.
 \item  In the US and Japan, both bearish and bullish indicators had low statistical significance, compared to the bearish being of high statistical significance and the bullish being of low statistical significance. 
 \end{itemize}This is interesting because it provides insight into the applicability of these indicators in more populated markets. This could help consumers understand which indicators not to use, as that is just as important as knowing which ones to use. The main takeaways can be summarized as follows for Heikin Ashi Candlesticks indicators:

\begin{itemize}
\item   
The bearish indicators were more significant compared to the bullish ones, which is consistent with our other results.
\item
For more populated housing markets such as the US and Japan, neither bullish nor bearish candlestick indicators were statistically significant. Hence,  these indicators are less likely to be significant with more volume in the market, as that proliferates any inherent randomness.

\end{itemize}

 \section{Concluding remarks}\label{last}
 
  In the present paper we have analyzed datasets provided by Zillow \cite{zillow} and FRED \cite{fred} for Japan, Germany, and Canada's historical housing prices. We then used technical indicators typically utilized in the Stock Market to identify trends with respect to our data.
    To determine the usability and frequency of technical indicators, we analyzed 4 candlestick patterns, we used the Wilcoxon test, developed in R code to calculate a $p$-value that signified statistical significance.     
    
\begin{figure}[H]
  \centering \hspace{-0.32 in} \includegraphics[scale = 0.48]{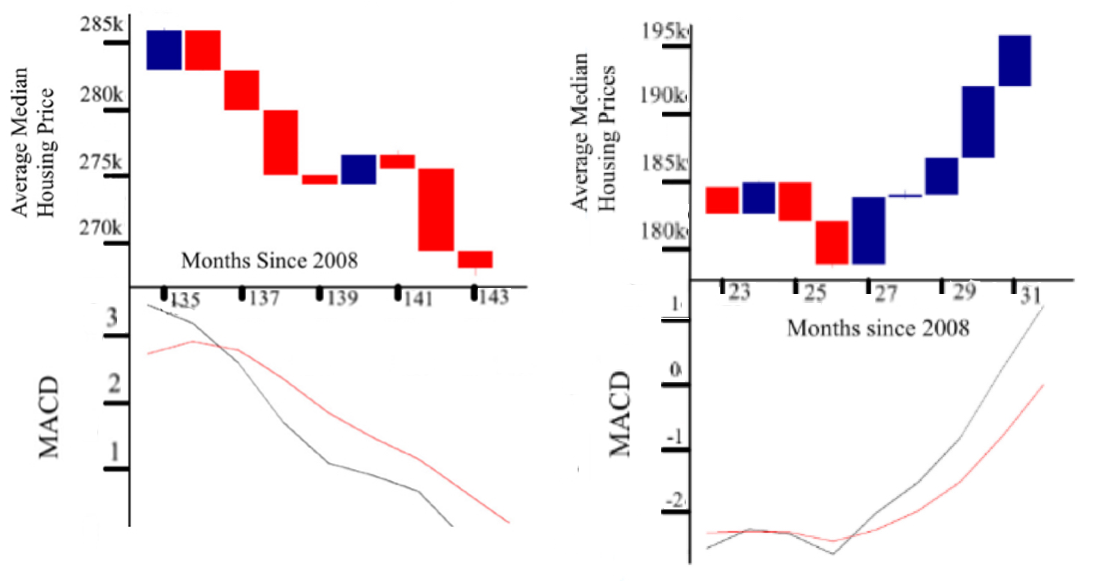}
  \caption{A visual demonstrating the predictive capacity of the MACD indicator, for both bearish and bullish trends.}
      \label{conclusionMACD}
\end{figure}

\begin{figure}[H]
  \centering\hspace{-0.3 in} \includegraphics[scale = 0.41]{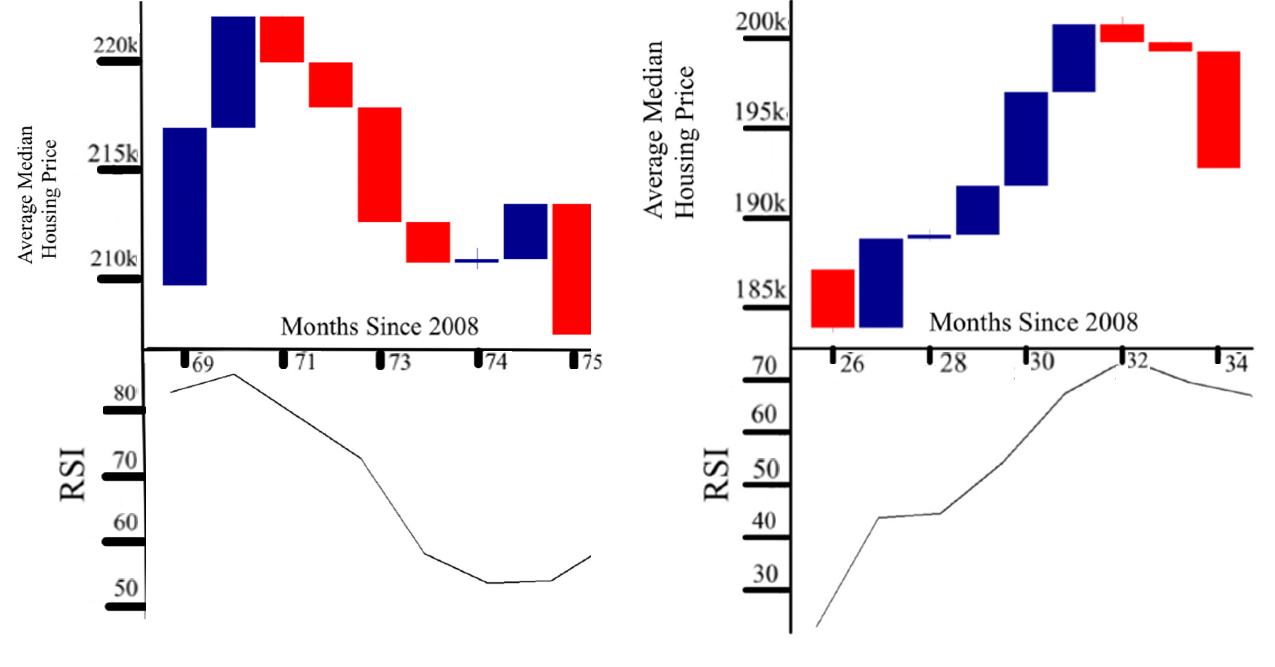}
  \caption{A visual demonstrating the predictive capacity of the RSI indicator, for both bearish and bullish trends.}
      \label{conclusionRSI}
\end{figure}

     An application of this research study is being able to reliably forecast future trends using these stock market indicators. \autoref{conclusionMACD} and \autoref{conclusionRSI} demonstrate the capacity of forecasting and its usability to understand housing market behaviours both within the US as well as for other global housing markets.       

   Through the study of  candlestick indicators via Heiken Ashi (Hammer, Hanging Man, Bearish Engulfing, Bullish Engulfing) we found the following behaviours: 
      \begin{itemize}
   \item[(I)] Populated/Saturated housing markets have low statisical significance values due to more randomness present in a more active market.\\
\item[(II)]  For the US housing market in particular, we found that bearish trends were less statistically significant compared to bullish trends.
\item[(III)] The converse of (I) and (II) is true when we examined the data for other types of markets through their representative countires, with bearish trends being less statistically significant compared to other countries. 
\item[(IV)] The difference in the significance of bearish and bullish trends varies depending on the market under consideration. In more volatile and saturated housing markets such as the US, Canada, and Japan, this difference is smaller as there is more randomness to decrease polarization, but in a stable market such as Germany this is not the case.
\end{itemize}

For nearly all of our analysis, RSI remained fairly strong in terms of its applicability. We conjecture this is the case because whilst most indicators analyze data over a large period of time, RSI places its emphasis on recent trends. Hence, a strong RSI lends itself to a bullish trend, and a weak RSI signifies a bearish trend.  In contrast, all of the other indicators such as MACD, Hanging Man, and Bullish Engulfing, an emphasis on later trends accounts for the decrease in statistical significance since there is a larger probability of external factors playing an important role. 
 
    Using the Wilcoxon test in R, we were able to determine the $p$-value of all of our indicators as shown in the table of \autoref{summaryTable1}. Through this study, one can deduce which indicators should be used in each type of market, and this is summarized in    \autoref{finalTable2}  below, where we present which indicators would be viable for various markets. This comes to embody the final results of our study, where we present the most effective indicators.

  \begin{figure}[H]
  \centering \includegraphics[scale = 0.45]{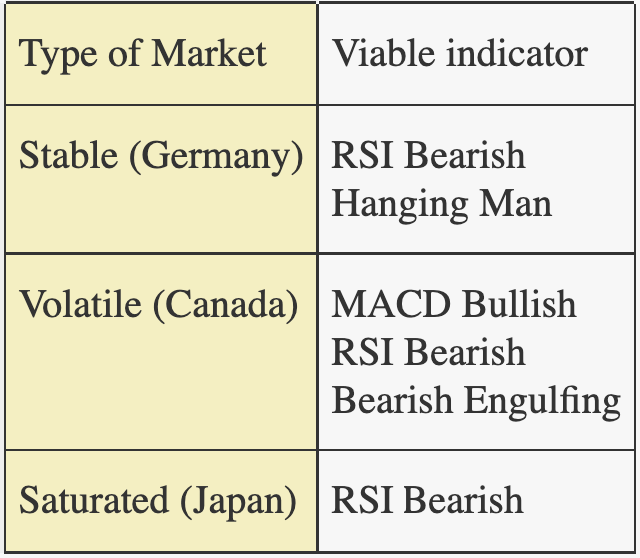}
  \caption{Indicators that can be used for the markets and countries   analyzed in our study, determined this by parsing through our $p$-values  in \autoref{summaryTable1} and determining the lowest $p$-values  per market and country, as these would yield the most statistically significant indicators for each case.}
      \label{finalTable2}
\end{figure}
It should be noted that the  RSI Bearish was  statistically significant across all markets, signaling that is an indicator that can be used in practically any case study. Moreover, from our study one can see that the difference in statistical significance between bearish and bullish indicators is smaller, as exemplified by Japan.

 In Japan, the difference between MACD Bearish and Bullish is 0.62, but in Germany is significantly larger, is 0.73. This means that in more crowded and volatile markets it can be more confusing and inherently variable about not only the success of statistically significant indicators but the failures of statistically insignificant indicators, whereas this notion is minimized as the difference gets larger in stable markets. This is because lower and higher $p$-values  dictate a more concrete sense of failure and success for statistically insignificant and significant indicators respectively.

The analysis of  6-month forecasts through the usage of MACD and RSI as shown in \autoref{conclusionMACD} and \autoref{conclusionRSI}, depict how consumers can forecast prices with these indicators to their benefit. Within the MACD analysis, observe that for the left images, consumers can deduce that for this period of 6 months (in this case Months 135-143 from 2008, so February to July 2020), prices will decrease, due to the fact the black MACD line crossed the red signal line from above. Similarly, for the right images consumers can deduce that for this period of 6 months (in this case Months 23-31 from 2008, so November 2009 to April 2010), prices will increase, due to the fact the black MACD line crossed the red signal line from below. This analysis can be  extended to the RSI indicator, where  one can  observe that from the left graph consumers can deduce that for this period of 6 months (in this case Months 69-75 from 2008, so August 2013 to March 2014), prices will decrease, due to the fact the RSI values crosses an upper threshold. Similarly, for the bottom visualization consumers can deduce that for this period of 6 months (in this case Months 26-34 from 2008, so April 2011 to September 2011), prices will increase, due to the fact the black MACD line crossed the red signal line from below.   

  In our study, we did an extensive analysis on utilizing stock market indicators upon housing data from a variety of housing markets, ranging from volatile to stable housing markets in Canada, Germany, Japan, and the US. An interesting future application is rentals/apartments, which could be of immense interest with the application of indicators because the renting market is even more volatile than the housing market  \cite{kim_lee_lee_2014}, since there is a plethora of other external factors being inherently present due to the rental being a temporary residence (for example, holiday season variables, or reasons for sublet, e.g., a number of college students take rentals when doing internships at a company headquartered at a city).

In the last decade, there has been significant innovation in stock market analysis and the evolution of the housing market. In the past, the stock market has been analyzed with less nuanced techniques such as intrinsic value and speculating human behavior \cite{CHAUVET200087}. However, this changed drastically with the rise of candlestick analysis, as well as the emergence of the MACD and RSI indicators \cite{bhansali2020trust}. Moreover, the onset of the COVID-19 pandemic has meant that the housing market has transformed significantly \cite{jrfm14030108}. This inspired us to present the application of stock market indicators on the housing market with data from a variety of housing markets, so people, no matter what type of market they are in, can  know which indicators are significant and how trends can be forecasted, which is salient in saving millions of dollars within the housing market \cite{ENGELHARDT1998135}.
 
Finally, it is important to note that bearish trends were less statistically significant in the US but more significant in other countries. It would be thus very interesting to analyze whether this would be true across other time series, such as the rental market described earlier. But in the meantime, it is significant to realize that there are certain stock market indicators that can forecast future trends in housing prices.     
    \bigskip

 
  \noindent {\bf  Contributions of authors.} The present work constituted Varun Mittal's 2022 research project under the supervision of Laura P. Schaposnik. The authors carried out the research and preparation of the present manuscript. The content on this manuscript is not intended to provide legal, financial or real estate advice. It is for information purposes only.   \\
 
  \noindent {\bf Funding.} The work of Laura P. Schaposnik is partially supported through the NSF grants  CAREER DMS 1749013, NSF FRG Award DMS- 2152107 and a Simons Fellowship. The work appearing here was finished whilst LPS was a RP at SLMath, and their support  is greatly acknowledged.   \\
 
 \noindent {\bf Affiliations.}\\
  (a) James B. Conant High School, Schaumburg, IL, USA\\
  (b)  University of Illinois, Chicago, IL, USA. 
  
   \bibliographystyle{fredrickson}
\bibliography{PRIMES2020}{

\begin{thebibliography}{CYX21}

\bibitem[BS20]{bhansali2020trust}
R.~Bhansali and L.~P. Schaposnik, ``A trust model for spreading gossip in
  social networks: a multi-type bootstrap percolation model,'' {\em Proceedings
  of the Royal Society A} {\bfseries 476} no.~2235, (2020) 20190826.

\bibitem[Car89]{doi:10.1080/00420988920080181}
D.~T. Carruthers, ``Housing market models and the regional housing system,''
  \href{http://dx.doi.org/10.1080/00420988920080181}{{\em Urban Studies}
  {\bfseries 26} no.~2, (1989) 214--222},
  \href{http://arxiv.org/abs/https://doi.org/10.1080/00420988920080181}{{\ttfamily
  https://doi.org/10.1080/00420988920080181}}.
 
\bibitem[CNL14]{jrfm7010001}
T.~T.-L. Chong, W.-K. Ng, and V.~K.-S. Liew, ``Revisiting the performance of
  macd and rsi oscillators,'' \href{http://dx.doi.org/10.3390/jrfm7010001}{{\em
  Journal of Risk and Financial Management} {\bfseries 7} no.~1, (2014) 1--12}.
  \url{https://www.mdpi.com/1911-8074/7/1/1}.

\bibitem[CP00]{CHAUVET200087}
M.~Chauvet and S.~Potter, ``Coincident and leading indicators of the stock
  market,''
  \href{http://dx.doi.org/https://doi.org/10.1016/S0927-5398(99)00015-8}{{\em
  Journal of Empirical Finance} {\bfseries 7} no.~1, (2000) 87--111}.
 
\bibitem[CYX21]{jrfm14030108}
K.~S. Cheung, C.~Y. Yiu, and C.~Xiong, ``Housing market in the time of
  pandemic: A price gradient analysis from the covid-19 epicentre in china,''
  \href{http://dx.doi.org/10.3390/jrfm14030108}{{\em Journal of Risk and
  Financial Management} {\bfseries 14} no.~3, (2021) }.
  \url{https://www.mdpi.com/1911-8074/14/3/108}.

\bibitem[EM98]{ENGELHARDT1998135}
G.~V. Engelhardt and C.~J. Mayer, ``Intergenerational transfers, borrowing
  constraints, and saving behavior: Evidence from the housing market,''
  \href{http://dx.doi.org/https://doi.org/10.1006/juec.1997.2064}{{\em Journal
  of Urban Economics} {\bfseries 44} no.~1, (1998) 135--157}.

\bibitem[fre]{fred}
``Housing.''
\newblock \url{https://fred.stlouisfed.org/categories/97}.

\bibitem[GLB19]{GONG201951}
C.~M. Gong, C.~Lizieri, and H.~X. Bao, ````smarter information, smarter
  consumers''? insights into the housing market,''
  \href{http://dx.doi.org/https://doi.org/10.1016/j.jbusres.2018.12.036}{{\em
  Journal of Business Research} {\bfseries 97} (2019) 51--64}.
t
\bibitem[Kha08]{6073000}
A.~Khalafallah, ``Neural network based model for predicting housing market
  performance,'' \href{http://dx.doi.org/10.1016/S1007-0214(08)70169-X}{{\em
  Tsinghua Science and Technology} {\bfseries 13} no.~S1, (2008) 325--328}.

\bibitem[Kir76]{10.2307/621308}
A.~M. Kirby, ``Housing market studies: A critical review,'' {\em Transactions
  of the Institute of British Geographers} {\bfseries 1} no.~1, (1976) 2--9.
  \url{http://www.jstor.org/stable/621308}.

\bibitem[KLL14]{kim_lee_lee_2014}
K.-H. Kim, C.-M. Lee, and Y.-M. Lee, ``Chapter 12: Rental housing system and
  housing market volatility: Monthly rent-based vs. asset-based systems,'' Feb,
  2014.
\newblock
  \url{https://www.elgaronline.com/view/edcoll/9781783472871/9781783472871.00020.xml}.

\bibitem[KM71]{doi:10.1068/a030243}
R.~M. Kirwan and D.~B. Martin, ``Some notes on housing market models for urban
  planning,'' \href{http://dx.doi.org/10.1068/a030243}{{\em Environment and
  Planning A: Economy and Space} {\bfseries 3} no.~3, (1971) 243--252},
  \href{http://arxiv.org/abs/https://doi.org/10.1068/a030243}{{\ttfamily
  https://doi.org/10.1068/a030243}}.  
\bibitem[LS21]{LIU2021110010}
S.~Liu and Y.~Su, ``The impact of the covid-19 pandemic on the demand for
  density: Evidence from the u.s. housing market,''
  \href{http://dx.doi.org/https://doi.org/10.1016/j.econlet.2021.110010}{{\em
  Economics Letters} {\bfseries 207} (2021) 110010}.

\bibitem[LU22]{liang_unwin_2021}
Y.~Liang and J.~Unwin, ``Covid-19 forecasts via stock market indicators,'' {\em
  Nature Scientific Reports} (2022) .

\bibitem[MST19]{https://doi.org/10.48550/arxiv.1909.10660}
D.~Matsunaga, T.~Suzumura, and T.~Takahashi, ``Exploring graph neural networks
  for stock market predictions with rolling window analysis,'' 2019.
\newblock \url{https://arxiv.org/abs/1909.10660}.

\bibitem[MS22]{varun121322_2022}
Mittal and Schaposnik's repository,
  ``Varun121322/housing-market-forecasts-via-stock-market-indicators: An
  analysis of housing market forecasts with popular stock market indicators,
  including macd, rsi, and candlestick indicators.,'' 2022.
\newblock
  \url{https://github.com/Varun121322/Housing-Market-Forecasts-Via-Stock-Market-Indicators}.

\bibitem[zil21]{zillow}
``Zillow data,'' Mar, 2021.
\newblock \url{https://www.zillow.com/research/data/}.

\end{thebibliography}
}

\end{document}